\documentclass{aa}
\usepackage{graphicx}
\usepackage{amsfonts,amssymb,latexsym}
\usepackage{psfig}

 \newcommand{\msun}{${\rm M}_{\odot}$}
 \newcommand{\fluor}{$^{19}{\rm F}$~}
 \newcommand{\kms}{${\rm km}~{\rm s}^{-1}$~}

\def\lesssim{\mathrel{\hbox{\rlap{\hbox{\lower4pt\hbox{$\sim$}}}\hbox{$<$}}}}
\def\gtrsim{\mathrel{\hbox{\rlap{\hbox{\lower4pt\hbox{$\sim$}}}\hbox{$>$}}}}

\def\lambdabar{\protect\@lambdabar}
\def\@lambdabar{%
\relax
\bgroup
\def\@tempa{\hbox{\raise.73\ht0
\hbox to0pt{\kern.25\wd0\vrule width.5\wd0
height.1pt depth.1pt\hss}\box0}}%
\mathchoice{\setbox0\hbox{$\displaystyle\lambda$}\@tempa}%
{\setbox0\hbox{$\textstyle\lambda$}\@tempa}%
{\setbox0\hbox{$\scriptstyle\lambda$}\@tempa}%
{\setbox0\hbox{$\scriptscriptstyle\lambda$}\@tempa}%
\egroup
}
\def\chem#1#2{$\rm{}^{#1}\kern-0.8pt#2$}
\def\reac#1#2#3#4#5#6{$\rm\,{}^{#1}\kern-0.8pt{#2}\,({#3}\,,{#4})\,
{}^{#5}\kern-0.8pt{#6}\,$}
\def\gsimeq{\,\,\raise0.14em\hbox{$>$}\kern-0.76em\lower0.28em\hbox
{$\sim$}\,\,}
\def\lsimeq{\,\,\raise0.14em\hbox{$<$}\kern-0.76em\lower0.28em\hbox
{$\sim$}\,\,}
\def\be{\begin{equation}} 
\def\ee{\end{equation}}
\def\beqy{\begin{eqnarray}}
\def\eeqy{\end{eqnarray}}
\def\bmlet{\begin{mathletters}}
\def\emlet{\end{mathletters}}

\begin{document}
\title{The thermonuclear production of $^{19}$F by Wolf-Rayet stars revisited}

   \author{A. Palacios\inst{1}, M. Arnould\inst{1} and G. Meynet\inst{2}}

   \offprints{M. Arnould}

   \institute{Institut d'Astronomie et d'Astrophysique, Universit\'e
              Libre de Bruxelles, CP 226, B-1050 Brussels, Belgium
         \and
               Geneva Observatory, 51 Ch. des Maillettes, CH-1290 Sauverny, Switzerland
               }

\date{Received --; accepted --}

\abstract{New models of rotating and non-rotating stars are computed for initial masses
between 25 and 120 M$_\odot$ and for metallicities $Z$ = 0.004, 0.008, 0.020 and 0.040 with the aim
of reexamining the wind contribution of Wolf-Rayet (WR) stars to the $^{19}$F enrichment of the interstellar
medium. Models with an initial rotation velocity $\upsilon_{\rm i}$ = 300
\kms  are found to globally eject less $^{19}$F than the non-rotating models.  We compare our new predictions
with those of Meynet \& Arnould (2000), and demonstrate that the $^{19}$F yields are very sensitive to the
still uncertain \reac{19}{F}{\alpha}{p}{22}{Ne} rate and to the adopted mass loss rates. Using the
recommended mass loss rate values that take into account the clumping of the WR wind and the NACRE reaction
rates when available, we obtain WR $^{19}$F yields that are significantly lower than predicted by Meynet \&
Arnould (2000), and that would make WR stars non-important contributors to the galactic $^{19}$F budget.
In view, however, of the large nuclear and mass loss rate uncertainties, we consider that the question of
the WR contribution to the galactic $^{19}$F remains quite largely open. 
}

\titlerunning{WR production of $^{19}$F}
\authorrunning{A. Palacios, M. Arnould \& G. Meynet}

\maketitle
%
\section{Introduction}
\label{intro}

The solar system has for long been the only location in the Universe with a known
fluorine (\chem{19}{F}) content. The very origin of this \chem{19}{F} has been a major
long-standing nucleosynthetic puzzle, in spite of the fact that it has the lowest solar
abundance among the nuclides ranging from carbon to calcium.

Since the beginning of the nineties, the situation has changed drastically, both
theoretically and observationally. The first quantitative prediction that \chem{19}{F} could
be produced thermonuclearly at a level compatible with the solar amount has been made by
Goriely et al. (1989). They identify a mode of production of
\chem{19}{F} that could develop in He-burning conditions following the reaction chains

\begin{eqnarray*}
                            &             &
(\beta^+)^{18}O(p,\alpha)^{15}N(\alpha,\gamma)^{19}F                \cr
                            & \nearrow    &  \hskip 4.2cm  \searrow            
                                  \cr 
^{14}N(\alpha,\gamma)^{18}F & \rightarrow &
(n,p)^{18}O(p,\alpha)^{15}N(\alpha,\gamma)^{19}F(\alpha,p)^{22}Ne.  \cr
                            & \searrow    &   \hskip 4.2cm  \nearrow           
                                  \cr
                            &             & \hskip 0.5cm
(n,\alpha)^{15}N(\alpha,\gamma)^{19}F                               \cr
\end{eqnarray*}

\noindent In this scenario, the synthesis of \fluor requires the availability 
of neutrons and/or protons. They are mainly produced by the reactions
\reac{13}{C}{\alpha}{n}{16}{O} and \reac{14}{N}{n}{p}{14}{C}.

The viability of the proposed He-burning \fluor production scenario has
been demonstrated in the framework of detailed models for aymptotic giant
branch (AGB) stars
(Forestini et al. 1992, Mowlavi et al. 1998, Lugaro et al. 2004), as well
as of massive stars evolving through the Wolf-Rayet (WR) stage (Meynet \&
Arnould 1993, 2000 hereafter Paper I; Stancliffe et al. \cite{STAN05}).  
Massive stars that do not
experience the WR phase are expected to produce instead an insignificant
amount of \fluor 
during their hydrostatic evolution
(Meynet \& Arnould 1993, Woosley \& Weaver 1995, Limongi
\& Chieffi 2003). Note that the Type II supernova explosions of massive
stars have also been claimed to be responsible for a \fluor production
through $\mu$- and $\tau$-neutrino spallation on \chem{20}{Ne}
(e.g. Woosley \& Weaver 1995). This production is highly uncertain, as it
is very sensitive to the poorly known neutrino energy spectra. It will not
be discussed here.

The most recent solar and meteoritic \fluor values are provided by Lodders
(2003) and Asplund et al. (2005). Observational efforts to determine \fluor
abundances outside the solar system have been largely triggered by the
early theoretical predictions of Goriely et al. (1989). In a companion
paper to the one of Forestini et al. (1992), Jorissen et al. (1992) provide
the first \fluor abundances measured in stars other than the Sun. They
analyse a number of s-processed enriched galactic MS, S, and N-type giants
having a near-solar metallicity. These observations demonstrate that AGB
stars are fluorine producers, nicely confirming the initial predictions by
Goriely et al. (1989) and Forestini et al. (1992). The derived \fluor
abundances correlate with the carbon and s-nuclide ones, a pattern that AGB
models can also account for, as first shown by Mowlavi et al. (1998). Very
recently, \fluor has also been detected in a sample of H-rich as well as
H-deficient PG1159-type hot post-AGB stars (Werner et
al. \cite{Werner2004}).  The H-rich stars, which are not especially C-rich,
show solar-like \fluor abundances. This is in line with the conclusion by
Jorissen et al. (1992) concerning the C-F correlation in AGB stars. In
contrast, the H-poor stars exhibit very large F overabundances ranging from
10 to 250 times solar. Werner et al.  (2004) suggest that this abundance
pattern might be explained by the operation of a late post-AGB shell flash
(see Herwig et al. 1999). As a complement, a F/H abundance ratio of $4.5
\times 10^{-8}$ has also been determined in a planetary nebula (Liu
1998). This value again agrees with the Jorissen et al. (1992) correlation
between F and C.

Fluorine data also exist for stars whose surface, in contrast to the
(post-)AGB stars, is not expected to be contaminated with the products of
their in-situ nucleosynthesis. This concerns a few near-solar metallicity K
and M giants also analysed by Jorissen et al. (1992). In addition, Cunha et
al. (2003) have studied a sample of red giant stars in the Large Magellanic
Cloud, as well as in the atypical galactic globular cluster $\omega$
Centauri. These low-metallicity giants exhibit sub-solar fluorine
  abundances (A(F)\footnote{A(X) = log[$n$(X)/$n$(H)]+12} $\in$ [3; 4.19])
, while both the K - M field giants and the three K - M pre-main
sequence low-mass stars of the Orion nebula cluster exhibit nearly solar
fluorine (A(F) $\simeq$ 4.55) (Cunha \& Smith 2005). This effect of
metallicity is in agreement with the general
behaviour of [F/O] exhibited by stars in different evolutionary phases
within the galactic disk.
Such observations of non-contaminated stars
spanning a range of metallicities are mandatory if one wants to build a
model for the evolution of the \fluor content of the Galaxy and of other
stellar systems (Renda et al. 2004). Finally, fluorine has also been
observed in various interstellar locations (Federman et al.  2004, and
references therein) in an attempt to constrain the \fluor nucleosynthesis
models.

The aim of this paper is to revisit the predictions of Paper I of the \fluor
yields of WR stars on grounds of new models for different masses and
metallicities, and to provide the first predictions of the \fluor production by
rotating WR stars.

\section{Input physics}
\label{inputphys}

The models used here are computed with the Geneva stellar evolution code from the Zero Age Main
Sequence (ZAMS) up to the end of the He-burning (HeB) phase and are listed
in Tab.~\ref{tabyields}. The physical ingredients, structural predictions,
and comparisons to observations are discussed at length by Meynet \& Maeder
(\cite{paperX}, Figs. 9-12 and Tabs. 1-2 ; \cite{paperXI}, Figs. 7-9 and
Tables 1 and 3). The main points of relevance to the \fluor synthesis are as follows:

\noindent (1)  The initial compositions for the different metallicities are selected as in Palacios et
al. (2005). The initial \fluor mass fraction $X_{19} \equiv X_{19}(Z,0)$ at metallicity $Z$ is derived from the
simple scaling $X_{19}(Z,0) = (Z/0.02)X_{19\odot}$, where 0.02 is the adopted metallicity 
for the Sun, and $X_{19\odot}$ = $4.1 \times 10^{-7}$ is the solar \fluor
mass fraction according to Grevesse \& Noels (\cite{GN93});

\noindent (2) The effect of rotation on the mass loss rate $\dot{M}$ is taken into account as in Maeder
\& Meynet (\cite{paperVII}). 
As reference $\dot{M}$, we adopt for the pre--WR stages the values proposed by Vink et al.
(\cite{V00}, \cite{V01}) who account for the occurrence of bi--stability limits which affect the wind
properties and mass loss rates.  Outside the domain covered by these authors, the rates from de Jager et
al. (\cite{deJager88}) are selected.  As the empirical $\dot{M}$ values are derived from stars with a
variety of rotation velocities, and as $\dot{M}$ decreases with these velocities, a reduction factor to the
empirical rates of 0.85 (Maeder \& Meynet \cite{paperVII}) is introduced for the non--rotating models. 
During the WR phase, we use the $\dot{M}$ prescriptions of Nugis \& Lamers (\cite{NL00}). These rates,
which account for the clumping of the winds, are 2 to 3 times smaller than the ones used in previous
non--rotating `enhanced mass loss rate' stellar models presented in Paper I. Note
that wind anisotropies induced by rotation are neglected. These anisotropies are indeed shown to be very
small for the initial velocity $\upsilon_{\rm i} = 300$ \kms (Meynet \& Maeder \cite{paperX}) selected in
this work (see point 6 below). This would not be true for higher initial velocities (Maeder \cite{Ma02});
 
\noindent (3) During the pre--WR phases, it is assumed that the mass loss rates have a 
metallicity dependence given by $\dot M(Z)=(Z/0.02)^{1/2} \dot M(0.02)$ (Kudritzki \& Puls
\cite{KP00}, Vink et al. \cite{V01}). In contrast, no metallicity dependence is introduced during the WR
stage;

\noindent (4) All the models are computed with moderate core overshoot.
The distance of overshoot is taken equal to $d = \alpha H_{\rm p}$, where
$H_{\rm p}$ is the pressure scale height at the Schwarzschild boundary and
$\alpha =0.1$. This value of $\alpha$ is twice as small as the value used
in the models of Paper I;

\noindent (5) The transport of the nuclides and of the angular momentum is described as in Maeder \&
Meynet (\cite{paperVII}) and Meynet \& Maeder (\cite{paperVIII}) ;  

\noindent (6) All the considered stars are assumed to rotate on the ZAMS at an initial
rate $\upsilon_{\rm i} = 300$ \kms. For $Z = 0.02$, this value leads to time averaged
equatorial velocities on the Main Sequence (MS) well in the observed range (between 200 and 250 \kms); 
\begin{table*}[t]
\caption{Values of $p_{19}^{\rm wind}(M_{\rm i},Z)$ in units of
  $10^{-6}$\msun~for the rotating and non--rotating
models also displayed in Fig.~\ref{yields19F} }
\begin{center}
 \begin{tabular}{c|c|cccc}
 \hline
\hline
 $M$(\msun) & $\upsilon_{\rm i}$ (\kms) & \multicolumn{4}{|c}{$p_{19}^{\rm
     wind}$ ($10^{-6}$ \msun) }\\ 
& & & & & \\
& & \footnotesize{Z} = 0.04 & \footnotesize{Z} = 0.02 & \footnotesize{Z} =
 0.008 & \footnotesize{Z} = 0.004\\
 \hline
120 & 300 & 5.38 & -24.13 & 18.6 & 29.1\\
120 & 0 & 51.86 & 110.05 & & \\
85 & 300 & 28.41 & & &\\
60 & 300 & 27.97 & 19.4 & -5.18 & -1.81\\
60 & 0 & 103.5 & 22.96 & & \\
40 & 300 & 27.67 & & -2.32 & -0.757\\
30 & 300 & & & -1.49 & -0.284\\
25 & 300 & 15.17 & -6.03 & &\\
25 & 0 & -1.6 & -0.201 & &\\
\hline
\end{tabular}
\end{center}
\label{tabyields}
\end{table*}

\noindent (7) The reaction rates adopted in Paper I have been updated by
the use of the NACRE data (Angulo et al. \cite{Ang99}), when available. The
rates of the reactions entering the chain displayed in Sect.~\ref{intro}
and that are not considered in NACRE are taken from the following
references: \reac{14}{N}{n}{p}{14}{C} (Brehm et al. \cite{Brehm}),
\reac{18}{F}{n}{p}{18}{O} (REACLIB, Thielemann et al. \cite{reaclib}),
\reac{18}{F}{n}{\alpha}{15}{N} (Caughlan \& Fowler \cite{CF88}, hereafter
CF88), and \reac{19}{F}{\alpha}{p}{22}{Ne} (CF88).

\section{The WR production of {\bf $^{19}$F}}
\label{WR_F19}

\begin{figure*}
\begin{center}
\includegraphics[height=15cm,angle=-90]{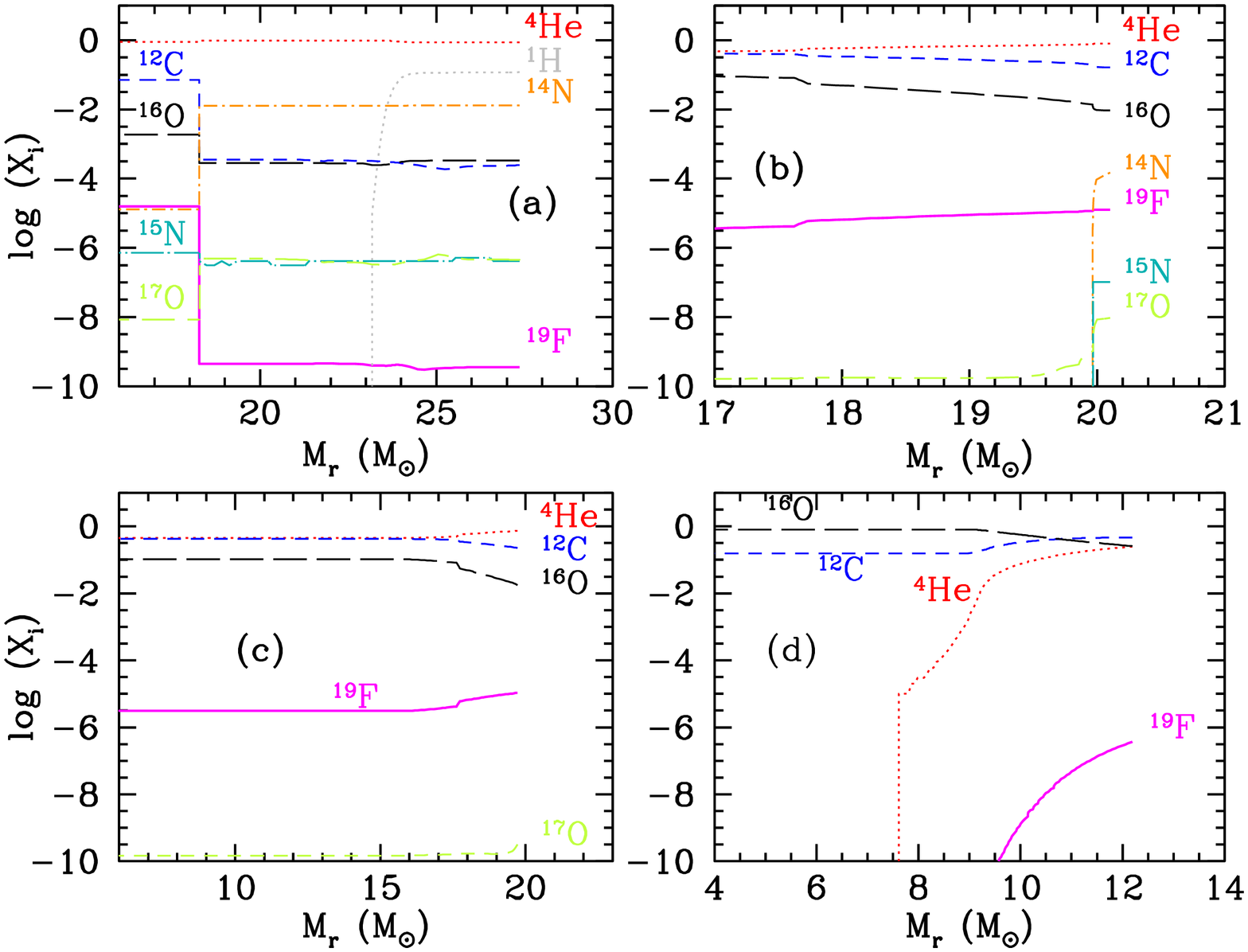}
\caption{Mass fractions of the CNO species and of \fluor versus mass {\it
M$_r$} (in M$_\odot$) inside a 60 M$_\odot$ with metallicity $Z = 0.02$ and $\upsilon_{\it
i}=0$ km s$^{-1}$ at three stages during the core He--burning phase (panels a--c), and at the beginning of
core C-burning (panel d). {\bf Panel {\bf (a)}: $X_{19}$ reaches its maximum value at the centre; {\bf (b)}:
$X_{19}$ starts to exceed about 10$^{-6}$ at the surface; {\bf (c)}: $X_{19}$ reaches its maximum value at
the surface; {\bf (d)}: There is no more fluorine in the core at the
beginning of the C-burning phase.}}
\label{figure1}
\end{center}
\end{figure*}

\begin{figure*}
\begin{center}
\includegraphics[height=15cm,angle=-90]{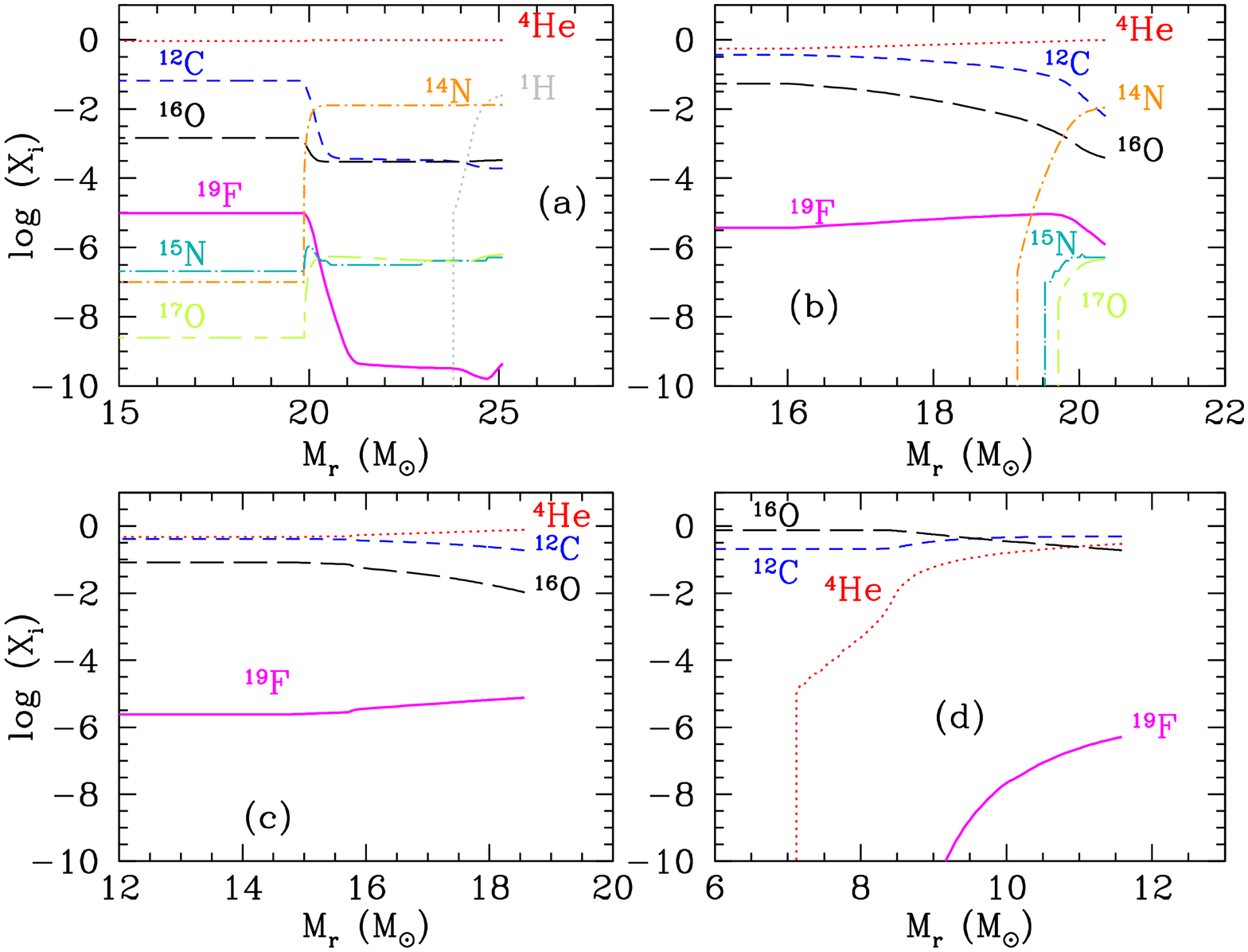}
\caption{Same as Fig.~\ref{figure1}, but for the $Z = 0.02$ 60 M$_\odot$ model with $\upsilon_{\rm i} =
300$~km s$^{-1}$.}
\label{figure2}
\end{center}
\end{figure*} 

Figure~\ref{figure1} shows the abundances of various nuclides inside the
non-rotating 60 \msun, $Z = 0.02$ model at three stages during the core
He-burning (HeB) phase associated with (a) the maximum mass fraction 
 $X_{19,\rm C}$ of fluorine at the centre, (b) the first time the mass
  fraction $X_{19,\rm S}$ of fluorine at the surface exceeds 10$^{-6}$, and
  {\bf (c) the maximum mass fraction $X_{19,\rm S}$ of fluorine at the
    surface. Panel (d) refers to the beginning of the C-burning phase, when
  there is no fluorine left in the core}. We focus on
the HeB stage since \chem{19}{F} is produced in the convective core at the
beginning of that phase mainly through the first chain of transformation of
\chem{14}{N} into \fluor displayed in Sect.~\ref{intro}. In only about 43
300 yrs, while the central He mass fraction $X_{4,\rm C}$ drops from 0.98
to about 0.90, the \fluor mass fraction at the centre $X_{19,\rm C}$ is
seen to increase from $4.4 \times 10^{-10}$ to its maximum value $1.6
\times 10^{-5}$ (see panel {\bf (a)}). At this time, the surface \fluor
mass fraction $X_{19,\rm S}$ is still very small ($3.6 \times
10^{-10}$). As evolution proceeds, the convective core retreats in mass,
leaving in its wake \chem{19}{F}-enriched layers that will eventually
appear at the surface when the star enters the WC phase (see panels {\bf
(b)} and {\bf (c)}). The strong wind during this phase finally allows the
ejection of these fluorine-rich layers in the interstellar medium
(hereafter ISM). Panel {\bf (b)} depicts the situation when $X_{19,\rm S}$
just exceeds 10$^{-6}$.  The time elapsed between panels {\bf (a)} and {\bf
(b)} is of the order of 148 000 yrs, corresponding to a $X_{4,\rm C}$ decrease
from 0.90 to 0.47. During this period, \fluor in the convective core is
partly destroyed by $^{19}$F($\alpha$,p)$^{22}$Ne. In panel {\bf (b)},
$X_{19,\rm C} = 3.7 \times 10^{-6}$, which is more than 4 times lower than
its maximum value. Concomitantly, mass loss starts exposing the most
\chem{19}{F}-rich layers at the surface (panel {\bf (c)}). From this stage
on, the synthesised \fluor starts enriching the ISM until the outer layers
are severely \chem{19}{F}-depleted, as it is the case at the beginning of
core C-burning (panel {\bf (d)}).

A generic sequence summarising the above can be used to describe the
evolution of the central and surface \fluor mass fractions in all the
models presented here. Fluorine is produced in the
convective core and its mass fraction rapidly increases at the beginning of
the He--burning phase. The regions above the convective core are not
\fluor-rich at this stage. As the evolution proceeds, the central
temperature becomes high enough for \reac{19}{F}{\alpha}{p}{22}{Ne} to
be efficient, leading to a decrease of the \fluor mass fraction in the convective
core. On the other hand, the convective core retreats in mass as a result of strong mass
loss. This allows part of the fluorine left behind by the
retreating core to escape destruction. This fluorine can then be exposed at
the surface by the stripping of the outer layers by stellar winds. Thus, at the surface, the \fluor mass fraction increases, reaches a
maximum, and then decreases as deeper layers are revealed.

From the above, it appears that WR stars can be \fluor
contributors to the ISM if at least :\\
 \indent (1) The star can enter the WC phase at a sufficiently early phase of
     core He-burning, so that the \fluor-enriched
   shells of the He--core, which coincide with those rich in carbon and oxygen, can
   appear at the stellar surface before fluorine is too much depleted (see Sect. 3).\\ 
\indent (2) The mass loss at the
beginning of the WC phase is high enough for removing efficiently the
\chem{19}{F}-rich layers. If mass loss is weak at this stage, part of the
\chem{19}{F} has time to be converted into \chem{22}{Ne} when deeper layers
are uncovered.\\ The fulfilment of these two conditions depends on initial
mass, metallicity and rotation velocity, as discussed below.

The \fluor ISM enrichment efficiency of a star with initial mass $M_{\rm i}$ and metallicity $Z$, undergoing
a wind phase of duration $\tau(M_{\rm i},Z)$ is conveniently evaluated in terms of its net \fluor yield
$p_{19}^{\rm wind}$ defined as 

\begin{eqnarray}
p_{19}^{\rm wind}(M_{\rm i},Z)  = \hspace{5.5cm} & \\ \nonumber
\int^{\tau(M_{\rm i},Z)}_0  [X_{19,\rm S}
  (M_{\rm i},Z,t) - X_{19,\rm S}(Z,0)]
 {\dot M} (M_{\rm i},Z,t)  dt . &
\end{eqnarray}

\noindent This quantity is a classical input in galactic chemical evolution
models. However, it does not represent the total yields of \fluor, since
more fluorine could be produced during the supernova explosion. Note also that
negative $p_{19}^{\rm wind}$ values are obtained when the ejected material contains less \fluor than
originally present in the star.  
The yields for the wind phase derived for both our rotating and
non--rotating models are displayed in Table~\ref{tabyields} in units of 10$^{-6}$~\msun.

\section{Sensitivity of {\bf $^{19}$}F production to initial conditions}
\label{sns_WR}
 
\subsection{Effect of rotation}

\begin{figure}
\begin{center}
\resizebox{\hsize}{!}{\includegraphics[width=8cm,height=8cm]{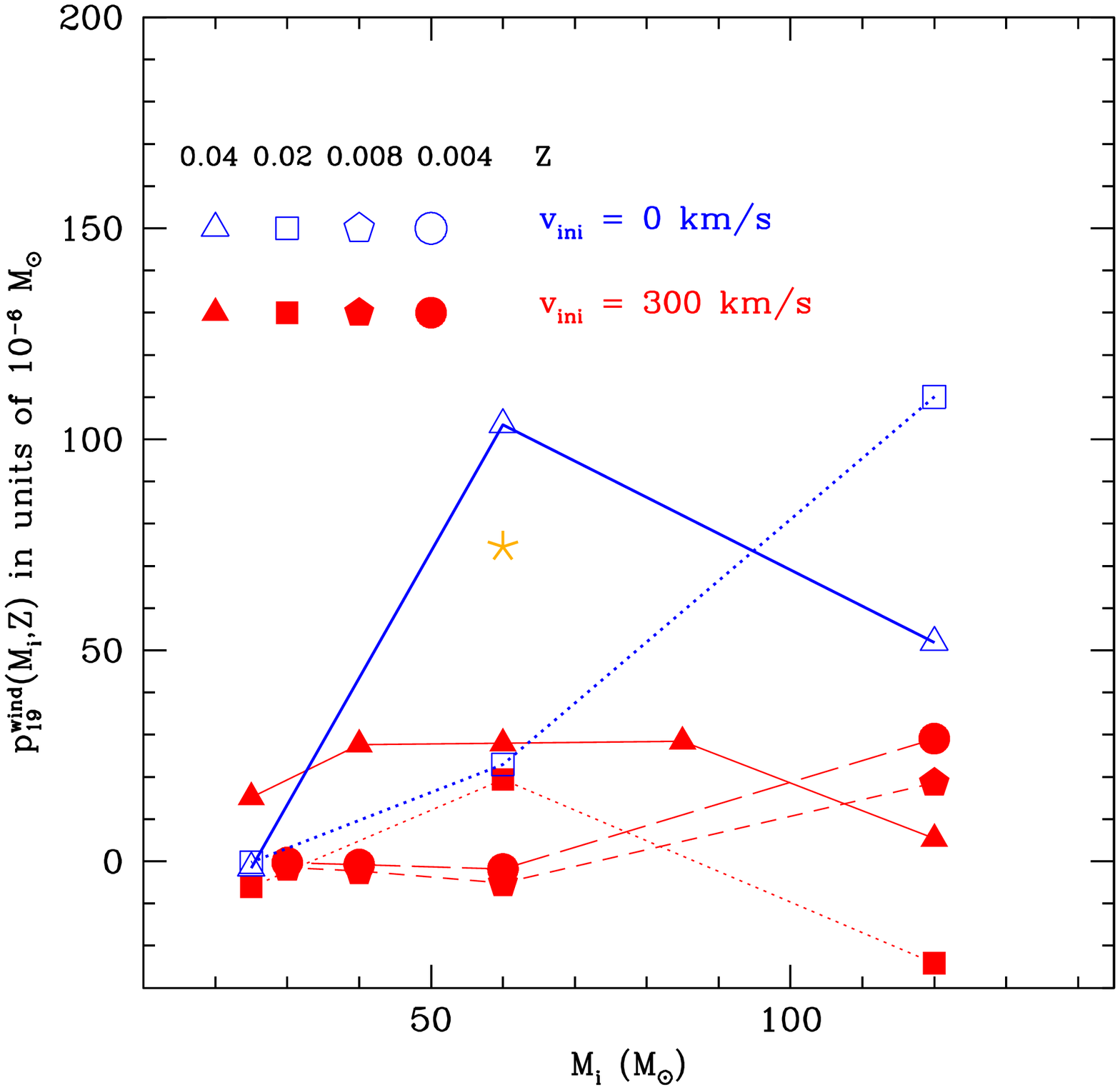}}
\caption{Values of $p_{19}^{\rm wind}(M_{\rm i},Z)$ in units of $10^{-6}$~\msun~for non-rotating (open
symbols) or rotating (filled symbols) models with $Z = 0.004$ to 0.04, the different metallicities being
represented as indicated on the figure. The black star represents $p_{19}^{\rm wind}(60,0.02)$ in the
non--rotating case, and with the CF88 rate for \reac{19}{F}{\alpha}{p}{22}{Ne} divided by 10.}
\label{yields19F}
\end{center}
\end{figure}

Our calculations indicate that rotation has a very limited impact on the
\fluor production in the stellar cores. This largely results from the fact
that rotation hardly affects the temperatures and densities in the core, as
well as from the inability of rotation to transport \chem{14}{N} from the
H-burning shell into the He core, which would significantly boost the
\fluor production. The insensitivity of the \fluor content to rotation for
the 60~\msun, $Z = 0.02$ model is seen from a comparison of
Figs.~\ref{figure1} and \ref{figure2}. Let us note that we cannot exclude
at this point the possibility of a \chem{14}{N} mixing into the core for
rotation velocities and metallicities different from the ones considered
here.

The limited role played by rotation on the \fluor content of the core does
not preclude changes in the \fluor yields, as shown in
Table~\ref{tabyields} and Fig.~\ref{yields19F}. In fact, $p_{19}^{\rm
wind}$ is found in all the considered cases to decrease when rotation is
included, except in the 25~\msun~case at $Z = 0.04$. The magnitude of this
reduction shows a high sensitivity to mass and metallicity. The
$p_{19}^{\rm wind}$ lowering with rotation can be explained as
follows. Rotation and associated transport processes favour an early
entrance of the stars into their WR phase (Fliegner \& Langer \cite{Fl95},
Meynet \& Maeder \cite{paperX}). Consequently, the period of high mass loss
rate is lengthened, with the result that the rotating models enter their WC
phase with, in general, a lower mass. At this stage, the mass loss rate
scales with the actual luminosity of the star (Nugis \& Lamers
\cite{NL00}), and thus with its actual mass, since the WC stars obey a
mass--luminosity relation (Schaerer \& Maeder \cite{SCMA92}). The
mass of the \chem{19}{F}-enriched material ejected into the ISM during the
WC phase is thus lowered as a result of rotation. This $p_{19}^{\rm wind}$ reduction appears
to be especially limited for the 60~\msun~model at $Z = 0.02$. This is due to a
balance between different effects acting in this particular case (see also
Meynet \& Maeder \cite{paperX}). The non--rotating model enters the WR
phase after a short Luminous Blue Variable (LBV) phase characterised by a
very high $\dot{M}$, while the WR phase with rotation already starts during
the MS, so that the LBV phase is skipped. The short LBV phase experienced
by the non-rotating star compensates for its later entry into the WR
phase. As a net result, the rotating and non--rotating 60 $M_\odot$ models
enter the WC phase with about the same mass. As far as the 
25~\msun~, $Z = 0.04$ case is concerned, the rotating star ejects some amount of \fluor
while its non-rotating counterpart does not. This derives from
the rotating model entering the WC phase, which is not the case in absence of
rotation.

In conclusion, mixing induced by rotation does not affect the amount of
\fluor synthesised in the central regions of the star at the beginning of
the core He-burning phase, at least for the range of metallicities explored
here. This might not be true any more at very low metallicity, in which
case shear mixing appears to be more efficient (Meynet \& Maeder
\cite{paperVIII}). Through structural effects (extent of convective cores,
mass-loss enhancement), rotation may however affect the overall evolution of
stars during the WR phase in such a way as to modify the $p_{19}^{\rm wind}$ values.

\subsection{Effect of mass and metallicity}
\label{metallicity}
  
\begin{figure}
\resizebox{\hsize}{!}{\includegraphics[width=8cm,height=8cm]{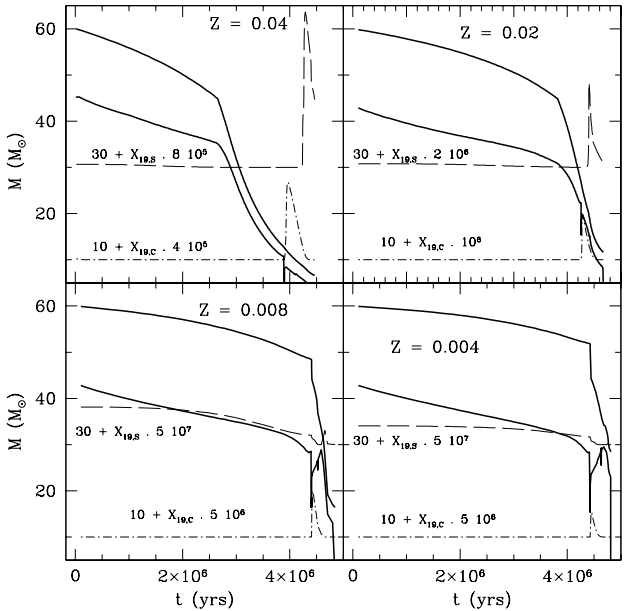}}
\caption{Evolution of $X_{19,\rm C}$ and $X_{19,\rm S}$ in a 60 \msun~star with
initial rotational velocity $\upsilon_{\rm i} = 300$ \kms at the four indicated metallicities. For
display purposes, $X_{19}$ is scaled appropriately. Solid lines describe the evolution of the total
and convective core masses.}
\label{60rotZ}
\end{figure}

Stars with different metallicities are likely to produce different amounts of \chem{19}{F}. This is confirmed
in Fig.~\ref{60rotZ} for a rotating 60 \msun~star. As in the non--rotating models, $X_{19,\rm C}$ increases
with metallicity (see Paper I). This relates directly to the enhanced production of $^{14}{\rm N}$ by the CNO cycles
during the central H--burning phase. The larger amount of \chem{14}{N} available at the beginning of the
HeB phase allows more \fluor to be synthesised. The very limited change of the central densities
and temperatures with metallicity does not affect this conclusion. Increased metallicities also favour higher 
$X_{19,\rm S}$ values as a result of larger mass loss
rates ($\dot{M} \propto Z^{1/2}$) which allow \chem{19}{F}-enriched layers to be exposed at the stellar
surface before the eventual partial transformation of \fluor by \reac{19}{F}{\alpha}{p}{22}{Ne}. In
contrast, some
\fluor destruction may be unavoidable at lower metallicities.  

These general considerations help interpreting the trends of the \fluor yields with mass and metallicity 
shown in Fig.~\ref{yields19F}. Let us just focus here on the models with rotation. For $Z < 0.02$,
$p_{19}^{\rm wind} < 0$ for $M_{\rm i} \leq 60$ \msun. The minimum mass for a rotating star to enter the WC
phase is $M_{\rm WC} \approx 25$ and 52 \msun~at $Z = 0.008$ and 0.004. Below these limits, the stars cannot
contribute to the \fluor enrichment of the ISM through winds. In the $M_{\rm WC} < M_{\rm i} \leq 60$
\msun~range, $X_{19,\rm S}$ has already dropped below its initial value by the time the He-burning products
appear at the surface, leading to negative $p_{19}^{\rm wind}$ values. In contrast, the $M_{\rm i} > 60$
\msun~stars contribute to the \fluor enrichment of the ISM because enough mass is lost in these cases for
uncovering the He core before the destruction of
\chem{19}{F}. For the 120 \msun~model, $p_{19}^{\rm wind}$ is larger at $Z = 0.004$ than at 0.008 because
the lower-$Z$ star enters the WC phase with a higher mass (48 instead of 31 $M_\odot$ for $Z$ increasing from
0.004 to 0.008), thus losing more mass at this stage.

The trend discussed above is reverted at $Z = 0.02$. In this case,
$p_{19}^{\rm wind}$ grows from negative values for the 25 \msun~star (at $Z
= 0.02$, this star only marginally enters the WR phase, and does not
contribute to the ISM \fluor enrichment) to a positive value for the 60
\msun~model. It then decreases again to a largely negative value at 120
\msun. For this particular case, the model enters the WR phase already
during the MS and loses more than 80 \% of its mass prior to the HeB
phase. These ejecta are not \fluor-enriched. When the wind becomes
\chem{19}{F}-rich at the beginning of the WC phase, the star has only a
small remaining mass and $\dot{M}$ is consequently small from this point
on, allowing only a limited amount of \chem{19}{F}-rich
material to be ejected. This results in a negative $p_{19}^{\rm wind}$ value.

At $Z = 0.04$, $p_{19}^{\rm wind}$ is positive for all models, and is seen to be rather insensitive to
stellar mass. This results from a subtle
balance between the dependence with mass of $X_{19,\rm S}$ and the amount of ejected \chem{19}{F}-enriched
material.

\subsection{Uncertainties in the $^{19}${\rm F} yields from $^{19}${\rm F}($\alpha$,\rm p)$^{22}$\rm
{Ne}}
\label{nuclear}

The \reac{19}{F}{\alpha}{p}{22}{Ne} reaction is the main \fluor destruction channel in the
considered stars (see Sect.~\ref{intro}). The large uncertainties on its rate of course 
concur with the problems of purely astrophysical nature to affect the reliability of the
predicted contribution of WR stars to the \fluor budget of the Galaxy. The status of our present
knowledge of the \reac{19}{F}{\alpha}{p}{22}{Ne} rate has been discussed recently by Lugaro et al.
(2004) and Stancliffe et al. (2005). The rate is still poorly known, the uncertainties
increasing dramatically with decreasing temperatures. The rate they recommend is more than one order of
magnitude smaller than the CF88 rate used in Paper I and in the present work. In
view of this, $p_{19}^{\rm wind}$ has been calculated for the 60 \msun, $Z = 0.02$ rotating model
decreasing the CF88 rate by a factor of 10. As a result, the yield is found to increase by
more than a factor of 3, as shown by the asterisk in Fig.~\ref{yields19F}. The yield could still be increased slightly by
decreasing the rate further. There appears to be room for this, as the lower limit
of the rate proposed by Stancliffe et al. (2005) is more than 14 orders of magnitude smaller than their
recommended rate at $T = 2 \times 10^8$ K! They however come to the conclusion that the corresponding yields
are only increased by at most 10 \% if this extremely small lower limit is adopted instead of their
recommended rate. This is not surprising, as the yields are essentially `frozen' as soon as the \fluor
lifetime against $\alpha$-captures becomes longer than its residence time in He-burning zones. This
situation is encountered if the CF88 rate for
\reac{19}{F}{\alpha}{p}{22}{Ne} is divided by 2000. Thus, for rates below this limit, $p_{19}^{\rm wind}$
becomes independent of the \chem{19}{F} $\alpha$-capture rates, and is just some percents higher
than those displayed in Table~\ref{tabyields} and Fig.~\ref{yields19F}. If \reac{19}{F}{\alpha}{p}{22}{Ne}
becomes small enough, one might wonder about other \fluor destruction channels, and in particular about the
precise role of its radiative neutron captures. In the 60 \msun~rotating star at $Z = 0.02$, the mass
fraction of neutrons is non-negligible only at the very centre of the star, but
decreases rapidly by several orders of magnitude further out in the convective core. Neutron captures are
thus not expected to be responsible for a significant destruction of \chem{19}{F}, even if the
$\alpha$-capture channel has a reduced efficiency.

\begin{figure}
\begin{center}
\includegraphics[width=8cm,height=8cm]{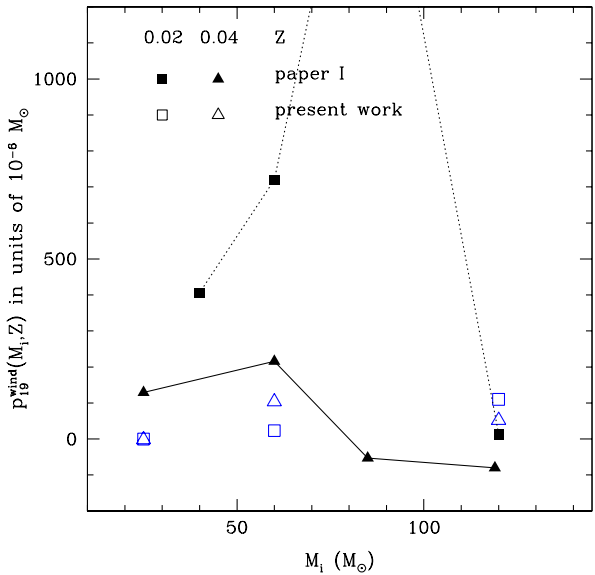}
\caption{Same as Fig.~\ref{yields19F} for the present
non-rotating models (open symbols) and from those of Paper I (filled symbols). Data obtained for $Z =
0.02$ and 0.04 are represented by squares and triangles.}
\label{compyieldsnorot}
\end{center}
\end{figure}

\section{Comparison with Paper I}

Figure~\ref{compyieldsnorot} is similar to Fig.~\ref{yields19F} and
presents the yields at $Z = 0.02$ and 0.04 obtained for the
non-rotating models of Paper I and of the present work. The main differences between the two sets of
computations lie in the nuclear reaction rates, the amplitude of the core overshoot, and the mass loss
prescriptions. For the relevant nuclear reactions, the rates adopted here are equal to, or differ only
marginally from the ones used in Paper I, except for
$^{15}$N($\alpha$,$\gamma$)$^{19}$F. The rate from CF88 used in Paper I is
replaced by the NACRE (Angulo et al. 1999) adopted one, which is about 34
times smaller than the CF88 one at $T = 2 \times 10^8$ K. However, this change does not affect the \fluor
production in the non-rotating models. At the beginning of He burning,
the $\alpha$-captures on \chem{15}{N} are very efficient even when adopting
the NACRE rate, and \chem{15}{N} is {\em completely} transformed
into \fluor well before fluorine starts turning into \chem{22}{Ne}. The net
amount of \fluor ultimately produced and ejected by WR stars is thus
controlled by the destruction channel \reac{19}{F}{\alpha}{p}{22}{Ne}
rather than by the production one.

Concerning the core overshoot its adopted value is lower in
  the present models than in those of \cite{paperI}. All other things being
  kept the same, we should expect the following consequence of this change. Lower core
  overshoot leads to smaller He-cores. Smaller He cores imply that more material
has to be removed for the core to be uncovered, which clearly disfavours \fluor
ejection by stellar winds. On the other hand, smaller central temperatures result.
This does not affect significantly the \chem{19}{F} production
from \chem{15}{N}, as \chem{15}{N} has anyway ample time to be totally transformed into \chem{19}{F}.
In contrast, the \chem{19}{F} destruction is slowed down, and this tends to make its yields greater.
Indeed more time is provided to stellar winds to remove the outer layers
before fluorine is destroyed in the core. Reduction of the amount of core
  overshoot should thus have two contrary effects whose relative importance
  remains unclear.\\ As already mentioned before, the standard models in \cite{paperI}
  not only differ from the present ones by the core overshoot, but by
  the adopted mass loss prescription as well, which also modifies the evolution of
  the core mass and temperature. 
 
Even if this additional difference
  prevents us from isolating clearly the effect of core overshoot, it
  appears however, that the differences with respect to \cite{paperI} are
  dominated by changes in the mass loss prescription. The prescription of Sect.~\ref{inputphys} reduces by a factor
of about 2 to 3 the rates adopted in \cite{paperI}. It is emphasised above that the revised prescription has to
be preferred, as it takes into account the clumping of the WR wind. In addition, it leads to a very good
agreement between the observed populations of O-type and WR stars and the predictions relying on
rotating models (Meynet \& Maeder \cite{paperX}, \cite{paperXI}). 

The impact of a change in the mass loss is intricate, and we just try here
to identify general trends. As already mentioned above, for a given star
to contribute to the \chem{19}{F} enrichment of the ISM, the mass loss
rates have to be large enough at the beginning of HeB to uncover the core
before the \fluor $\alpha$-particle captures become efficient enough. From
this, one might expect that $p_{19}^{\rm wind}$ increases with increasing
mass loss rates at the WC phase. However, as $\dot{M}$ scales with the
actual mass during this phase, the removal efficiency of the
$^{19}$F-rich layers is larger for more massive WC stars, which thus need
to have lost a relatively small mass during the previous evolutionary
phases. In summary, the WR $p_{19}^{\rm wind}$ yields depend drastically on
the mass loss prescriptions. They are large only if the mass loss rates are
high enough for removing most of the outer layers at the very beginning of
the HeB phase, but low enough for the star to keep a relatively high mass when it
enters the WC phase.

The general considerations developed above are confirmed by a closer
analysis of the non--rotating 60 and 120 M$_\odot$ models. As shown in
Fig.~ \ref{compyieldsnorot}, the former star is an extreme illustration of
the cases for which the present models predict p$_{19}^{\rm wind}$ values
lower than those reported in Paper I, with a reduction larger than a factor
of about 30. Even if, as made plausible by the discussion above, the
differences in the predicted yields likely result from the combined (and
difficult to disentangle) effects of the various changes in the ingredients
of the two sets of stellar models, the revised mass loss rates are most
probably responsible for the new situation encountered for the
60 \msun~model. The Paper I 60 \msun~ model at $Z = 0.02$ indeed enters the WC
phase with a mass of about 24 \msun~and $X_{4,\rm C} = 0.79$. Only less
than 5 \msun~remain at the end of the evolution. The newly computed
model enters the WC phase with a mass of about 21 \msun. This mass is
close to the one computed in Paper I, but is reached later in the
evolution, at a point where $X_{4,\rm C} = 0.47$. This delay is of course
the direct result of the lower mass loss rates adopted here in the previous
phase. In addition, the final mass of the star is 12.4 \msun. In other words, only 8.6
\msun~is lost during the WC phase, which is less than half of what is
computed in Paper I. All these effects tend to reduce $p_{19}^{\rm wind}$.

In contrast to the situation characterising the $M \leq 60$\msun~models,
the new \fluor yields for the non--rotating 120\msun~model are larger than
those of Paper I (Fig. \ref{compyieldsnorot}). In Paper I, the star is predicted to lose so much mass that it
enters the WC phase with only about 6 M$_\odot$. Just a small amount of
material can thus be lost during this phase (as $\dot{M} \propto M$ at this
phase). This clearly prevents any large \fluor yield. With the new
lower $\dot{M}$ values, the same star enters the WC phase with a mass
larger than 43 M$_\odot$, and more than 27 M$_\odot$ are lost during the WC
phase. This favours higher \fluor yields.
   
\section{Conclusions}
\label{conclusions}

Revised \fluor yields from non-rotating WR stars and the first evaluation of the yields from rotating such
stars are presented. The new yields in absence of rotation are significantly
lower than those of Paper I, as illustrated by the $Z = 0.02$ 60 M$_\odot$ case, where the reduction
amounts to more than a factor of 30. Rotation does not help in this matter, and even reduces the
yields. This drastic decrease of the predicted \fluor yields mainly results from the adoption of
reduced mass loss rates, and to a lesser extent from the selection of a smaller core overshoot.
 
Taken at face, these new predictions discard WR stars as important sources of the galactic
\chem{19}{F}. Let us however emphasise that they suffer from uncertainties originating from at least
two sources. As discussed above, $p_{19}^{\rm wind}$ is very sensitive to (1) the still
poorly known \reac{19}{F}{\alpha}{p}{22}{Ne} rate (see also Stancliffe et al. \cite{STAN05}), (2) the
still uncertain wind mass loss rates. Interestingly, Meynet \& Maeder (\cite{paperXI})
suggest on the basis of arguments concerning the WR population at solar and higher than solar metallicities,
that the mass loss rates during the post-core H-burning WNL phase might be underestimated. Higher mass loss
rates during this short stage would uncover more rapidly the He-core and would likely favour the ejection of
$^{19}$F.  

All in all, we consider that the question of the contribution of WR stars to the galactic $^{19}$F
remains largely open. It appears reasonable at this point to refrain from drawing any far-reaching
conclusion based on the present WR yield predictions, particularly in attempts to build galactic chemical
evolution models.

\acknowledgements
AP acknowledges financial support from the ESA
PRODEX contract 96009.

\end{document}